\documentclass[pdflatex,sn-mathphys-num]{sn-jnl}
\usepackage[utf8]{inputenc}
\usepackage{pdfcomment}
\usepackage{graphicx}%
\usepackage{multirow}%
\usepackage{multicol}
\usepackage{amsmath,amssymb,amsfonts}%
\usepackage{amsthm}%
\usepackage{mathrsfs}%
\usepackage[title]{appendix}%
\usepackage{xcolor}%
\usepackage{textcomp}%
\usepackage{manyfoot}%
\usepackage{booktabs}%
\usepackage{algorithm}%
\usepackage{algorithmicx}%
\usepackage{algpseudocode}%
\usepackage{listings}%
\usepackage{hyperref}
\usepackage{siunitx}%
\usepackage{upgreek}
\usepackage{comment}
\usepackage{svg}
\usepackage{makecell}
\usepackage{subcaption}
\begin{document}
 
\title{Optimization studies of silicon remoTES cryogenic calorimeters}
\author[1]{G.~Angloher}
\author[1]{M.R.~Bharadwaj}
\author[2,3]{A.~B\"ohmer}
\author*[1,2,3]{\sur{S.~Braun}}\email{sarah.braun@oeaw.ac.at}
\author[2,3]{M.R.~Cababie}
\author[5]{I.~Colantoni}
\author[4,5]{I.~Dafinei}
\author[4,6]{N.~Di~Marco}
\author[1]{C.~Dittmar}
\author[4]{F.~Ferella}
\author[4,5]{F.~Ferroni}
\author[2]{S.~Fichtinger}
\author[7,6]{A.~Filipponi}
\author[2]{M.~Friedl}
\author[2,3]{D.~Fuchs}
\author[8]{L.~Gai}
\author[1]{M.~Gapp}
\author[9]{M.~Heikinheimo}
\author[1,12]{K.~Heim}
\author[9]{K.~Huitu}
\author[2,3]{M.~Kellermann}
\author[2,3]{R.~Maji}
\author[1]{M.~Mancuso}
\author[4,6]{L.~Pagnanini}
\author[1]{F.~Petricca}
\author[4]{S.~Pirro}
\author[1]{F.~Pr\"obst}
\author[7,6]{G.~Profeta}
\author[6]{A.~Puiu}
\author[2,3]{F.~Reindl}
\author[1,12]{K.~Sch\"affner}
\author[2,3]{J.~Schieck}
\author[2,3]{P.~Schreiner}
\author[2,3]{C.~Schwertner}
\author[7,6]{P.~Settembri}
\author*[1]{\fnm{} \sur{K.~Shera}}\email{kshera@mpp.mpg.de}

\author[1]{M.~Stahlberg}
\author[9]{A.~Stendahl}
\author[6,10]{M.~Stukel}
\author[6,11]{C.~Tresca}
\author[8]{S.~Yue}
\author*[1,2]{\sur{V.~Zema}}\email{vanessa.zema@mpp.mpg.de}
\author[8]{Y.~Zhu}
\author[1]{L.~Ziegele}
\author[9]{N.~Zimmermann}


\affil[1]{\orgname{Max-Planck-Institut f\"ur Physik}, \city{Garching}, \postcode{85748}, \country{Germany}}

\affil[2]{\orgname{Marietta-Blau-Institut für Teilchenphysik}, \city{Wien}, \postcode{1010}, \country{Austria}}

\affil[3]{\orgname{Atominstitut, Technische Universit\"at Wien}, \city{Wien}, \postcode{1020}, \country {Austria}}

\affil[4]{\orgname{Gran Sasso Science Institute},  \city{L'Aquila}, \postcode{67100}, \country{Italy}}

\affil[5]{\orgname{INFN - Sezione di Roma}, \city{Roma}, \postcode{00185}, \country{Italy}}

\affil[6]{\orgname{INFN - Laboratori Nazionali del Gran Sasso}, \city{Assergi}, \postcode{67100}, \country{Italy}} 

\affil[7]{\orgname{Dipartimento di Scienze Fisiche e Chimiche, Universit\`a degli Studi dell'Aquila}, \city{L'Aquila}, \postcode{67100}, \country{Italy}}

\affil[8]{\orgname{SICCAS - Shanghai Institute of Ceramics}, \city{Shanghai}, \postcode{200050}, \country{P.R.China}}

\affil[9]{\orgname{Helsinki Institute of Physics, Univ. of Helsinki}, \city{Helsinki}, \postcode{00014}, \country{Finland}}

\affil[10]{\orgname{SNOLAB}, \city{ P3Y 1N2 Livel}, \postcode{1010}, \country{Canada}}

\affil[11]{\orgname{CNR-SPIN c/o Dipartimento di Scienze Fisiche e Chimiche, Universit\'a degli studi dell'Aquila}, \city{L'Aquila}, \postcode{67100}, \country{Italy}}
\affil[12]{\orgname{Technische Universit\"at M\"unchen}, \city{M\"unchen}, \postcode{80333}, \country{Germany}}

\abstract{The remoTES design, developed within the COSINUS experiment, enables a broader range of materials to be operated as cryogenic calorimeters read out with Transition Edge Sensors (TESs). In this configuration, the TES is fabricated onto a separate chip and thermally coupled to the absorber via a gold (Au) link. The remoTES concept has been successfully tested on various target materials. To further enhance detector performance and to fully exploit the advantages of the remote coupling design a series of optimization studies has been conducted using silicon (Si) absorbers as benchmark. This work presents an evaluation of several measurements aimed at reducing the thermal boundary resistance and enhancing signal transmission across Si remoTES interfaces, specifically from the absorber to the phonon collector and from the phonon collector to the TES. By testing a new Au link design and three distinct phonon collector configurations, Au, copper, and aluminum (Al)/Au, we achieved a baseline resolution of (21.5±0.3) eV using the Al/Au phonon collector.}
 
\maketitle
 
\section{Introduction}\label{sec1}
 
TESs are ultra-sensitive sensors deposited on absorbers to detect the energy deposited by particle interactions. They are operated at the transition between the superconducting and normal-conducting states of a thin-film material. In this regime, even a tiny temperature increase caused by a particle interaction, on the order of \si{\micro\kelvin}, induces a measurable change in resistance of \si{\milli\ohm}. This sharp superconducting transition enables TESs to detect extremely small energy depositions with excellent energy resolution in the sub-eV to eV range \cite{CRESST:2024cpr, BILLARD2024116465}, enabling the detection of rare, low-energy events. TES-based detectors have been successfully deployed in direct detection dark matter (DM) and neutrino experiments, including CRESST-III \cite{Abdelhameed_2019, CRESST:2024cpr}, SuperCDMS \cite{SuperCDMS:2020aus, SuperCDMS:2024yiv}, TESSERACT \cite{chang2025limitslightdarkmatter}, and NUCLEUS \cite{NUCLEUS:2025ymr}, as well as to X-ray astronomy, such as the X-IFU instrument on ESA's Athena observatory~\cite{Barret2018XIFU}.

Standard tungsten (W) TES deposition processes involve elevated temperatures and wet chemistry. Extending this technology to additional target materials, such as NaI, the target of the COSINUS experiment, is challenging, since NaI is soft, hygroscopic, and has a low melting point making it incompatible with these processes. To address this, the COSINUS collaboration developed the remoTES design, in which the TES is fabricated on a separate chip and coupled to the absorber via a Au link~\cite{Angloher_2023}.
This separation offers two key advantages. First, it enables the use of absorber materials otherwise incompatible with standard TES fabrication. Access to a broader range of absorber materials is beneficial for a variety of physics searches, as different target nuclei offer complementary sensitivity to different interaction types and particle candidates, extending beyond dark matter direct detection to other rare-event searches~\cite{Cerde_o_2013}. Second, fabricating the TES on a separate, dedicated chip allows the fabrication process to be optimized independently of the absorber, improving the reproducibility of TES production, a key requirement for next-generation, large-scale experiments~\cite{artuso2013sensorcompendium}. However, to fully profit from these advantages, we need to demonstrate that the remoTES can achieve a performance comparable to that of standard TES. 

In this work, we present recent optimization studies of the remoTES design using Si absorbers. The strategy is to enhance baseline resolution by improving signal transmission across the remoTES interfaces, achieved through optimizing two design parameters: the geometry of the Au link on the TES chip, which we call the Au port (see Figure~\ref{remoTES_of_COSINUS}), and the choice of phonon collector material deposited on the absorber. Testing and reducing the thermal boundary resistance between detector components is the driving motivation behind the optimization, as it governs the efficiency of the signal transmission across each interface. Si is chosen as a benchmark material because it is widely used as a cryogenic calorimeter absorber with the standard TES, making it well suited for direct comparison.

The manuscript is structured as follows: Section~\ref{sec:design} introduces the general remoTES design. Section~\ref{sec:theory} covers the formation of the signal from a particle interaction and the signal transmission across the relevant thermal boundaries. Section~\ref{sec:results} reports the measurement and analysis results, and Section~\ref{sec:concl} presents the conclusions and outlook.
 
\section{The remoTES design}
\label{sec:design}
\begin{figure}
    \centering
   \includegraphics[width=0.60\linewidth]{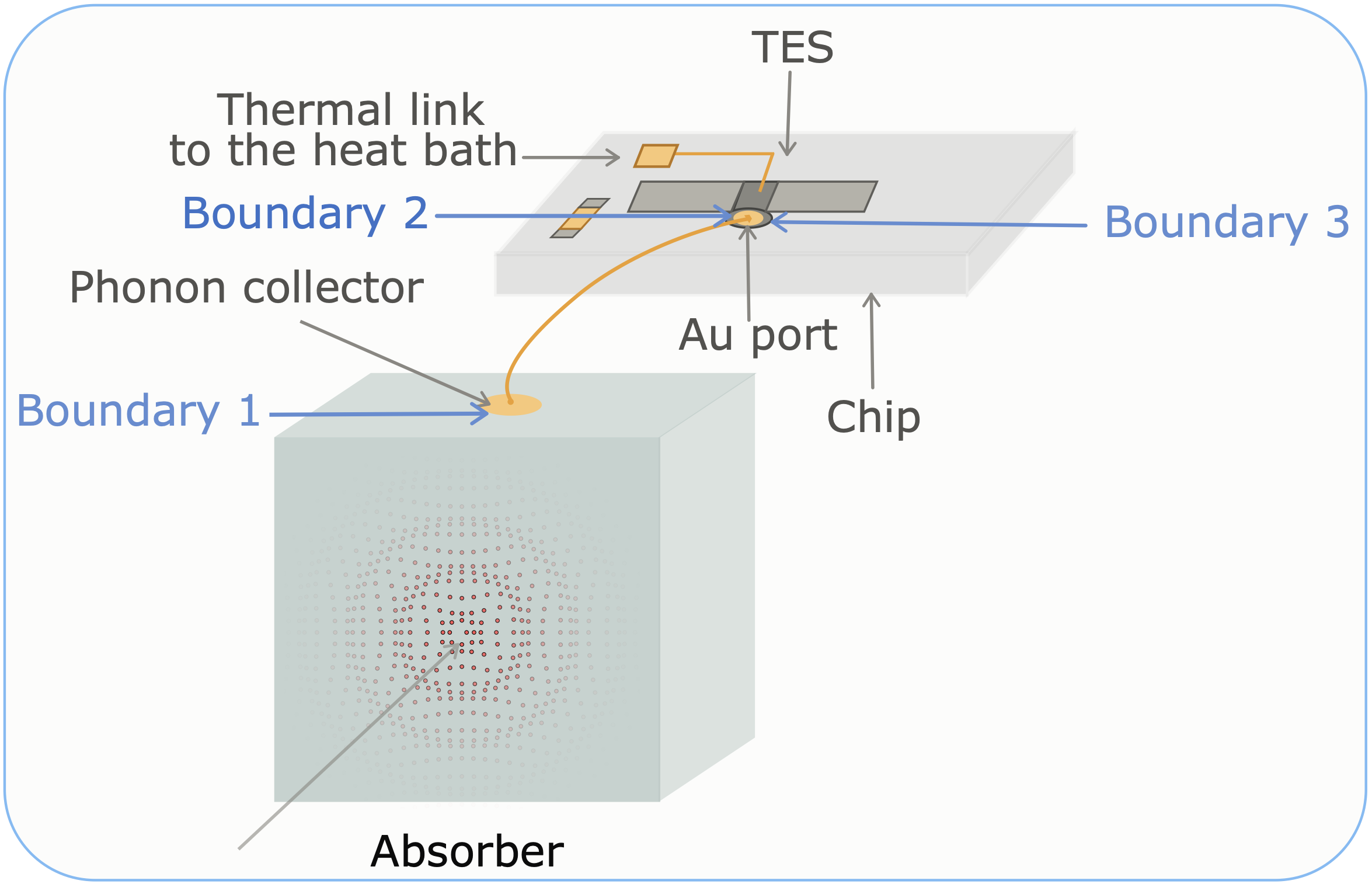}
    \includegraphics[width=0.39\linewidth]{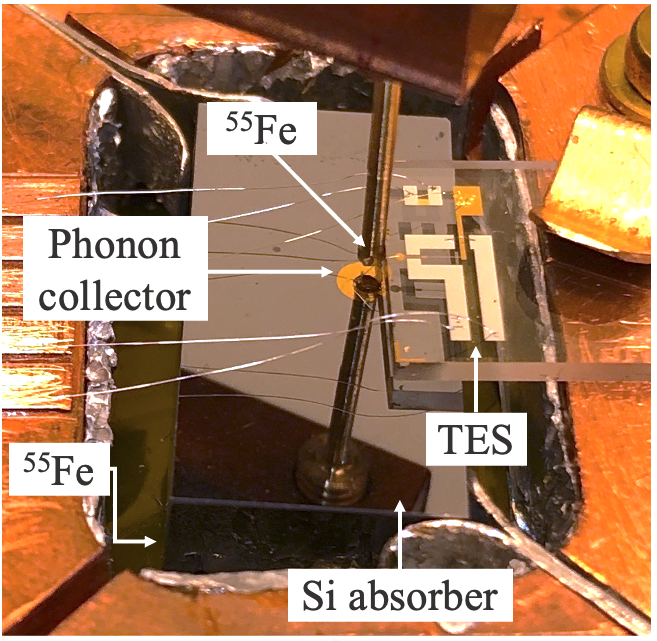}
    \caption{{Left: Schematic representation of the remoTES design implemented in the COSINUS experiment. The thermal boundaries considered in this work are indicated in blue. Further details are provided in the main text. Right: An image of a Si remoTES detector with the Au phonon collector.}}
    \label{remoTES_of_COSINUS}
\end{figure}

In the remoTES design, depicted in Figure~\ref{remoTES_of_COSINUS}, the TES is deposited on a separate substrate, a sapphire (Al$_2$O$_3$) chip in this work, and thermally coupled to the absorber via an intermediate link \cite{Angloher_2023, article, PhysRevD.110.043010, angloher2025cosinusmodelindependentchallenge}.
The link consists of three elements: a phonon collector deposited directly onto the surface of the absorber; a Au port, a Au film fabricated on the TES chip; and a Au bonding wire connecting the phonon collector to the Au port, providing the thermal link between absorber and TES. Phonons generated in the absorber following a particle interaction propagate to the phonon collector, where they are absorbed. The collection mechanism depends on the phonon collector material. In a normal-conducting collector, phonons couple to the electronic system via electron–phonon coupling~\cite{Angloher_2023}. In a superconducting/normal-conducting phonon collector, phonons with energy above the pair-breaking energy (h$\nu$$>$$2\Delta$)~\cite{Yen2014} instead break Cooper pairs in the superconducting layer, producing quasiparticles that diffuse into the adjacent normal-conducting layer, where they thermalize. The heat then propagates through the electronic system of the Au wire and reaches the TES, where it heats the TES's electronic system, generating a measurable signal.

This indirect coupling scheme offers several key advantages. From a fabrication and scalability perspective, (i) it preserves the radiopurity of the absorber, since the absorber is not exposed to TES fabrication processes; (ii) it is compatible with a broad range of materials, including hygroscopic or fragile absorbers such as NaI, which previously could not be employed as cryogenic calorimeters; and (iii) it enables scalable and reproducible TES fabrication on dedicated chips, facilitating wafer-level mass production for large-scale detector arrays. From a detector development perspective, (iv) it allows for systematic studies in which a single TES chip can be paired with different absorber materials of varying masses and geometries; and (v) it permits the investigation of alternative phonon collector materials and geometries. In this work, we exploit the latter two advantages to systematically optimize the performance of Si remoTES detectors.

\section{Thermal signal transmission}
\label{sec:theory}
 
\subsection{Signal formation}
The deposition of energy from an incoming particle interacting in a Si absorber  occurs via excitation of phonons and electron-hole pair formation~\cite{knoll2010radiation}. Since our sensors are only sensitive to phonons, we focus on the energy transported by phonons. The initial population of high frequency phonons, each of $\mathcal{O}(10)$~THz, has a short decay time of the order of picoseconds~\cite{PhysRev.148.845,PhysRevB.50.14179}. This population decays into lower energy phonons via three-phonon down-conversion processes~\cite{srivastava2022physics}. The decay cascade stops when phonons are fully thermalized with the rest of the system.
We refer to athermal phonons as the non-thermal phonon population produced promptly after the interaction, which undergoes successive anharmonic decay and scattering while propagating through the absorber. Only after sufficient down-conversion and multiple scattering events does the phonon distribution approach equilibrium; we refer to this population as thermal phonons, characterized by the lattice temperature.
 
\subsection{Signal transmission and collection}
\label{sub:thermalboundaryresistance}
In a Si remoTES detector with a Au phonon collector, the signal is transmitted from the Si absorber through the TES, across several thermal boundaries: Si absorber/Au phonon collector $\rightarrow$ Au phonon collector/Au wire $\rightarrow$ Au wire/Au port $\rightarrow$ Au port/W film $\rightarrow$ W film/Au thermal link to the bath.
In this work, we focus on reducing the thermal resistance across three boundaries specific to the remoTES implementation:
\begin{enumerate}
\item Si absorber/Au phonon collector
\item Au wire/Au port
\item Au port/W film.
\end{enumerate}
The thermal boundary resistance to the phonon transmission from the absorber to the phonon collector (boundary 1) is determined by the quality of the interface (for example area in contact, boundary layers, lattice defects at the interface created during sensor fabrication, possible cracks), and the matching of the phonon dispersion curves across the interface~\cite{PhysRev.60.354, probst1995model,RevModPhys.61.605, PhysRevD.110.112007,khot2025phonon}. It corresponds to the inverse of the Kapitza conductance, which we indicate as $G_K$. The following phonon absorption in the electronic system of the phonon collector depends on the strength of the electron-phonon coupling in the phonon collector and the corresponding conductance $G_{ep}$~\cite{probst1995model, RevModPhys.61.605, tong2019comprehensive, PhysRevD.110.112007,PhysRevB.101.201404}. In this work, we measured three Si-remoTES, each with a different phonon collector material (Au, Cu and a superconducting/normal-conducting phonon collector of Al/Au) to optimize the transmission through boundary 1 and the following collection of the signal. Al is a widely used phonon collector material~\cite{Abdelhameed_2019, chang2025limitslightdarkmatter, NUCLEUS:2025ymr} because of its higher critical temperature of about \SI{1.2}{\kelvin}~\cite{cochran1958superconducting} with respect to W, thus negligible heat capacity at millikelvin~\cite{pobell2007matter}. Additionally, a larger electron-phonon coupling than in Au and Cu is expected~\cite{PhysRevB.100.144306}. We envisioned to combine it with Au to profit from the quasiparticle trapping mechanism~\cite{irwin1995quasiparticle}, in combination with the requirement of coupling the phonon collector to the Au bonding wire in the remoTES design.

\begin{figure}
    \centering
    \includegraphics[width=0.65\linewidth]{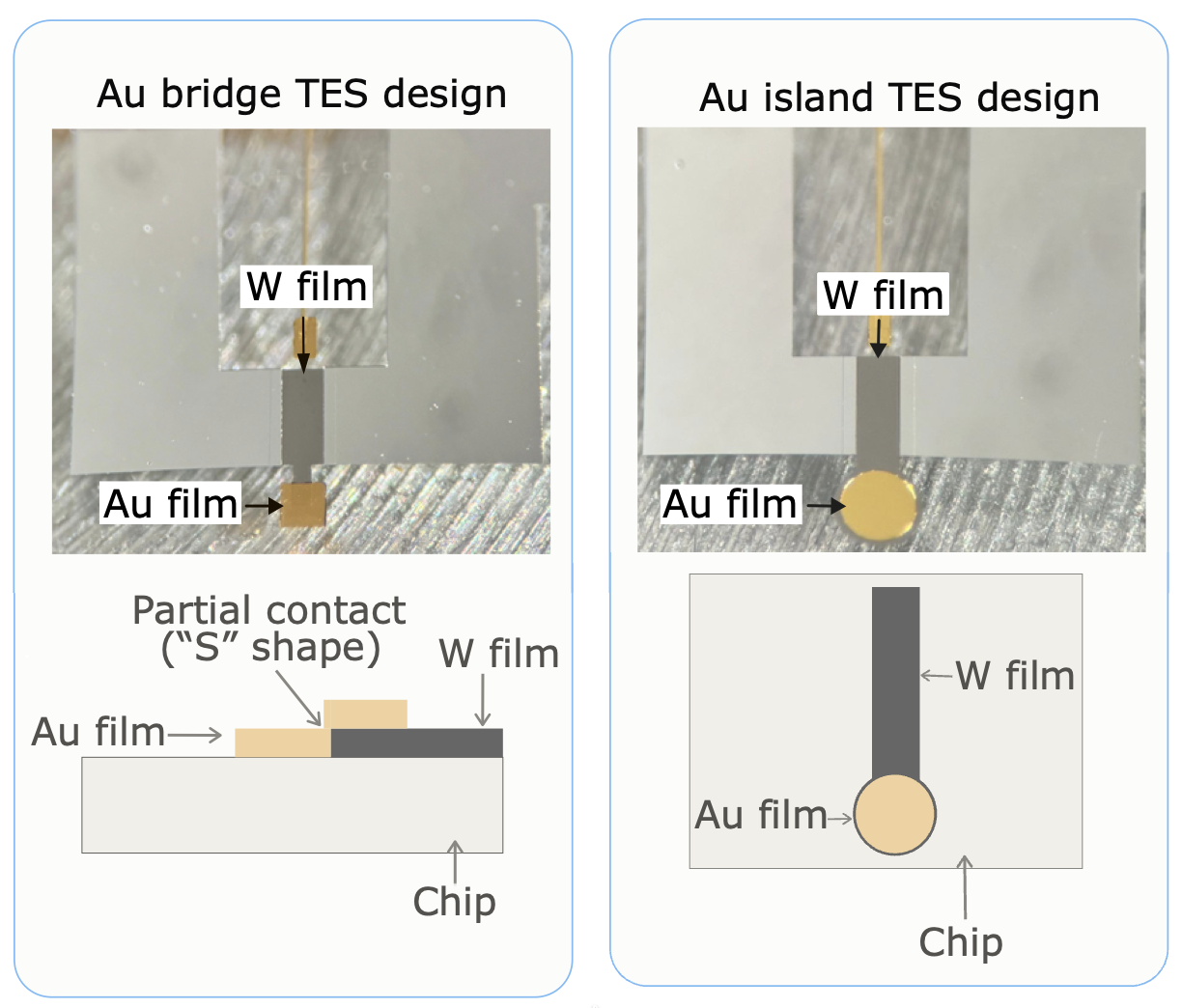}
    \caption{Left: Image of the Au bridge TES design (top) and its cross-sectional view (bottom). Right: Image of the Au island TES design (top) and top view illustration (bottom). Further details in the text.}
    \label{islandvsbridge}
\end{figure}

 The optimization of the conductance across the boundaries 2 and 3 (Au wire/Au port and Au port/W film) consists of a design change of the Au port, from an ``S" shape in contact with both the chip and the W film (dubbed Au bridge TES), to a disk shape entirely on top of the TES W film (dubbed Au island TES), with no contact to the chip, to avoid losses and improve the conductance from the Au port to the TES W film. Images of the two different designs, as well as an illustration of the “S” shape, are shown in Figure~\ref{islandvsbridge}. We reported on the first measurements with this design in~\cite{article}\footnote{Another thermal boundary which could affect the thermal transport is the type of foot of the bonding wire connecting the phonon-collector to the Au port. Studies comparing wedge bonding and ball bonding were previously performed on Si remoTES detectors with glued \cite{article, key} and sputtered phonon collectors \cite{key}. No significant improvement in detector performance was observed. However, ball bonding was established as the standard method because it is less destructive to the phonon collector and to the absorber itself.}. We follow up in this work with a validation of those results.\\
 
\section{remoTES optimization studies}
\label{sec:results}
To enhance the performance of the remoTES, a series of systematic optimization studies were previously conducted \cite{article,key,dissertationMK,dissertationFW}. In the present work, we report on additional measurements that validate and extend these initial investigations.
 
\subsection{Signal transmission from the phonon collector to the TES}
\label{sub:AubridgeAuisland}
 The TES design fabricated on a Al$_2$O$_3$ chip incorporates a Au port, which serves both as the bonding pad and as a contact to the underlying W film. In the original configuration (Au bridge TES) the Au port was partially on the W film and partially extended onto the Al$_2$O$_3$ chip, forming a direct interface with the chip (see Figure~\ref{islandvsbridge}, left panels). The partial contact with both the W film and the chip, creating the "S" shape indicated in Figure~\ref{islandvsbridge}, was considered a possible bottleneck for the signal transmission: the thermal signal arriving at the Au port must travel in-plane along the Au pad before entering the W film through a comparatively small contact area, resulting in a relatively low thermal conductance. To address this bottleneck, the design was revised so that the Au port is deposited entirely on the W film, resulting in a relatively high contact area between Au and W. This modified design is referred to as the Au island TES (see Figure~\ref{islandvsbridge}, right panels).
 
 Previous studies have shown that the Au island configuration results in a faster pulse decay compared to the Au bridge design, for events occurring both in the absorber and within the Au phonon collector itself \cite{article, key}. This finding was further validated by a dedicated measurement using a (20$\times$10$\times$5)~mm$^3$ Si absorber with a circular sputtered Au phonon collector (thickness: \SI{0.2}{\micro\meter}; area: \SI{3}{\milli\meter\squared}) and a Au island TES design. A $^{55}$Fe source was positioned beneath the absorber to irradiate the absorber, while an additional $^{55}$Fe source was collimated directly onto the Au phonon collector. The connection between the Au phonon collector on the absorber and the Au port on the TES was realized with a 25 \si{\micro\meter} Au bonding wire using ball bonding on the absorber side and wedge bonding on the TES side. 
 
\begin{figure}
    \centering
    \includegraphics[width=0.45\linewidth]{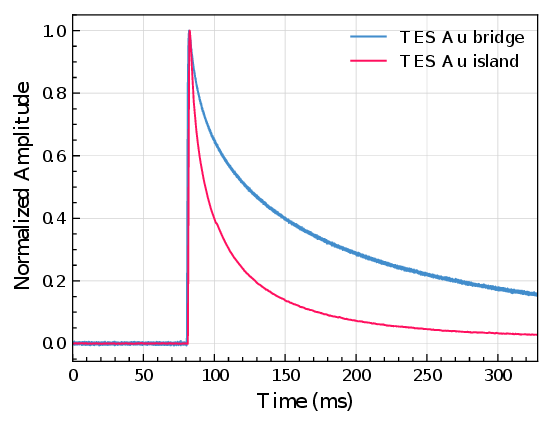}
    \includegraphics[width=0.45\linewidth]{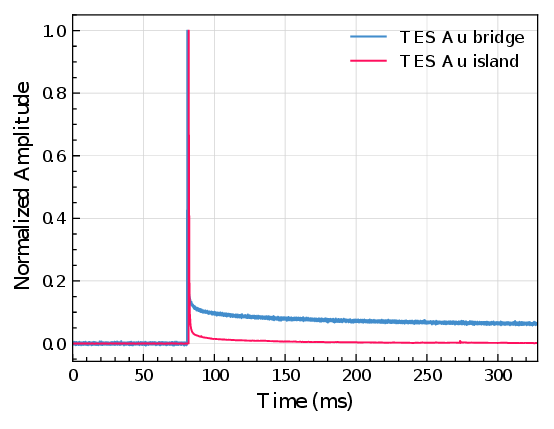}
    \caption{Pulse shapes of events in the absorber (left panel) and in the Au phonon collector (right panel), measured using TESs with the Au bridge (blue) and the Au island (pink) designs.}
    \label{Pulse shapes Au bridge vs Au island}
\end{figure}

 Figure~\ref{Pulse shapes Au bridge vs Au island} presents the standard event (SEV), defined as the normalized average pulse shape for events with the same energy deposition \cite{Angloher_2023}, shown separately for direct hits in the absorber (left panel) and for direct hits in the Au phonon collector (right panel). A comparison between the SEV of the Au bridge (blue) and Au island (pink) detector designs shows that the Au island design consistently yields faster-decaying pulses. Two competing effects govern the thermal transport across the Au port/W interface in these two designs. On one hand, the full contact between the Au port and the W pad in the island design increases the thermal conductance across this interface relative to the bridge design, where only a small W tip is contacted, this favors faster decay. On the other hand, the heat capacity of the W film coupled to the Au port is higher, as the W pad now is much bigger than in the Au bridge design. Since the Au island design nonetheless shows faster pulse decay overall, we conclude that the increase in thermal conductance dominates over the accompanying increase in heat capacity in determining the overall decay time. Based on these results, the Au island TES has been adopted as the standard remoTES design for COSINUS and will be implemented in the NaI remoTES detectors for the first COSINUS physics run at Laboratori Nazionali del Gran Sasso.
 
From a series of optimization studies, the best-performing Si remoTES detector with a Au phonon collector achieved a baseline resolution of \SI{89(2)}{\electronvolt} \cite{key}. Notably, significantly better performance has been demonstrated using Si absorbers with the TES deposited directly onto the absorber surface: the CRESST and TESSERACT collaborations have reported baseline resolutions of \SI{1.36(0.05)}{\electronvolt}~\cite{CRESST:2022lqw} and \SI{361.5(0.4)}{\milli\electronvolt}~\cite{chang2025limitslightdarkmatter}, respectively. Although the two detector designs differ in heat capacity as well as in transition temperature and shape, the observed large discrepancy suggests that phonon collection efficiency at the absorber/Au phonon collector interface may be a limiting factor, as discussed in Sec.~\ref{sub:thermalboundaryresistance}. To address this, a new series of optimization studies focusing on the signal transmission across the absorber/phonon collector thermal boundary and following signal collection is presented below.
 
\subsection{Signal transmission from the absorber to the phonon collector}
\label{sec:measurment}
\begin{figure}
    \centering
   \includegraphics[width=0.95\linewidth]{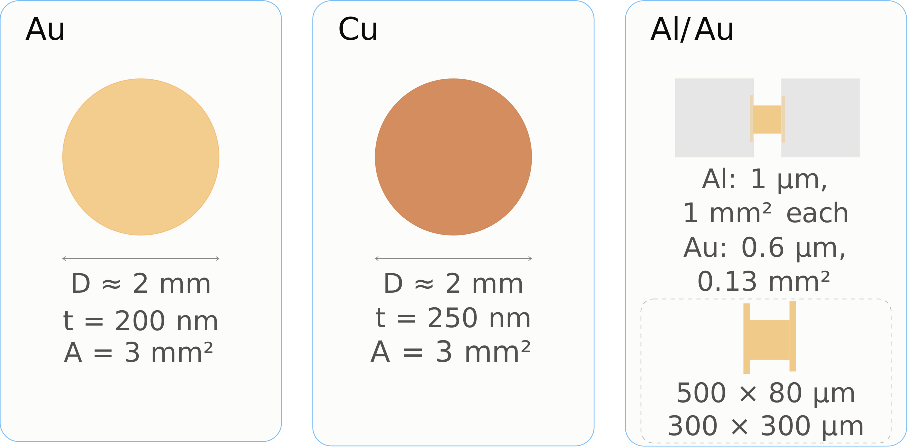}

    \caption{{Schematic representation of the different phonon collector designs used in this manuscript.}}
    \label{phononcollectors}
\end{figure}

To investigate the effect of phonon collector material on the signal transmission from the absorber to the phonon collector and consequently on the baseline resolution, three detector configurations were studied, each employing a different phonon collector material: Au, Cu, and Al/Au, as described in Section~\ref{sub:thermalboundaryresistance}. The technical specifications of each detector design are summarized in Table~\ref{baselineresolution}. The Au and Cu phonon collectors each have a circular geometry with an area of \SI{3}{\milli\meter\squared}. The Al/Au phonon collector consists of two Al pads, each with an area of \SI{1}{\milli\meter\squared}, and one Au pad with an area of \SI{0.13}{\milli\meter\squared}, with an Al/Au overlap area of \SI{0.04}{\milli\meter\squared}. A schematic representation of the phonon collector designs used in the Si remoTES detectors is shown in Figure~\ref{phononcollectors}.

To ensure a controlled and systematic comparison, only the phonon collector was varied across the three detectors, while all other components, including the TES, bonding technique, $^{55}$Fe  calibration sources, and detector holder, were held identical. The bonding wire length was also kept comparable across all three detectors, so that the only significant difference in module heat capacity arises from the phonon collector itself, whose values for each design are listed in Table~\ref{baselineresolution}. A picture of the assembled Si remoTES detector with the Au phonon collector is shown in Figure~\ref{remoTES_of_COSINUS}.

\begin{figure}
    \centering
    \includegraphics[width=0.45\linewidth]{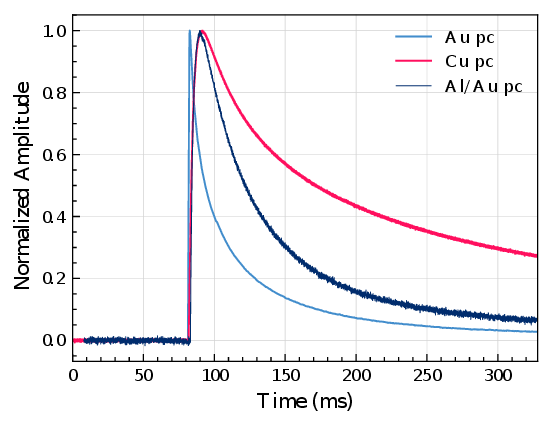}
    \includegraphics[width=0.46\linewidth]{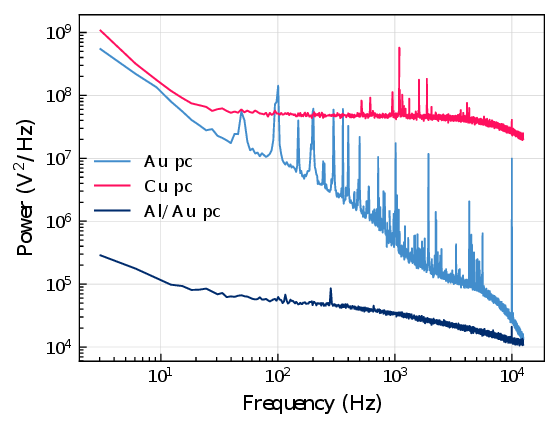}
    \caption{Left panel: Normalized standard event templates (SEVs) for
             absorber events for the three Si~remoTES detector
             configurations studied in this work, each equipped with a
             different phonon collector material. Right panel: Corresponding noise power spectral densities in units of \SI{}{\electronvolt\squared\per\hertz}, calibrated using the 5.89 keV Mn K$_\alpha$ line from the $^{55}$Fe source. In both panels, the
             detectors equipped with the Au, Cu, and Al/Au phonon
             collectors are shown in light blue, pink, and dark blue,
             respectively. }
    \label{Au_Cu_Al/Au_phonon_collector}
\end{figure}

The SEVs for absorber events, together with the noise power spectra for each detector configuration, are presented in Figure~\ref{Au_Cu_Al/Au_phonon_collector}. For direct absorber hits, the detector equipped with the Cu phonon collector exhibits a slower rise time and a significantly longer decay time compared to the Au phonon collector configuration, while the detector with the Al/Au phonon collector displays a rise time comparable to that of the Cu configuration and a decay time lying between those of the Au and Cu configurations. The slower rise and decay times observed for the Cu phonon collector, relative to Au, can be attributed to its higher heat capacity and weaker electron-phonon coupling compared to Au~\cite{Tong:2019phtherm}. In contrast, the slower pulse shape observed for the Al/Au configuration relative to Au cannot be explained by heat capacity, since the Al/Au phonon collector has a lower heat capacity than the Au phonon collector alone. Instead, this difference may be attributed to the fundamentally different signal formation mechanisms in Al and Au: rather than coupling directly to the electronic system via electron-phonon coupling as in Au, phonons in Al break Cooper pairs, producing quasiparticles that must subsequently diffuse across the Al film before reaching the Au pad. This additional diffusion step, absent in the Au-only design, may therefore introduce an extra timescale into the signal formation process~\cite{Yen2014}, plausibly contributing to the slower pulse response observed in the Al/Au configuration.

\begin{figure}[htbt!]
    \centering
    \includegraphics[width=0.30\linewidth]{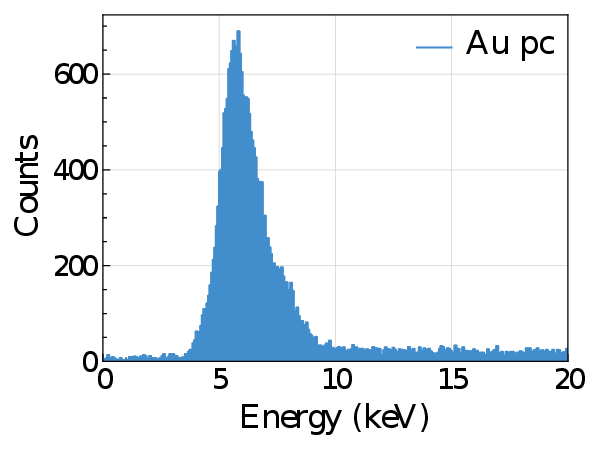}
    \includegraphics[width=0.30\linewidth]{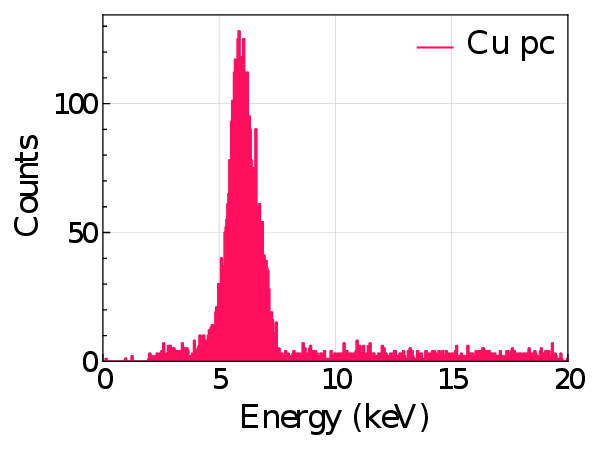}
    \includegraphics[width=0.30\linewidth]{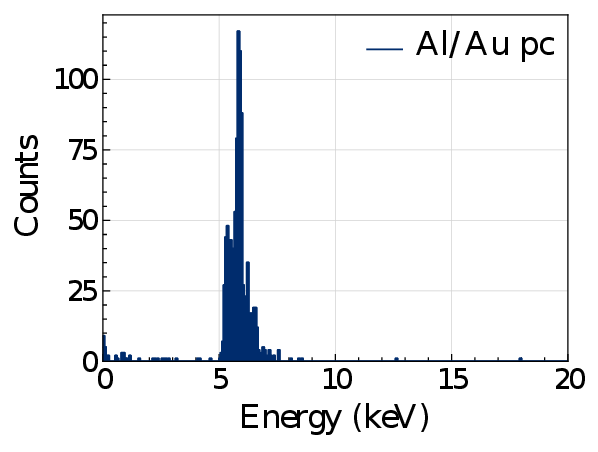}
    \caption{Calibrated energy spectra of absorber events for the
             three Si~remoTES detector configurations, each acquired with different exposure times. Left panel: Detector
             with Au phonon collector (light blue). Middle panel: Detector
             with Cu phonon collector (pink). Right panel: Detector with
             Al/Au phonon collector (dark blue).}
    \label{Spectra_Au_Cu_Al/Au}
\end{figure}

In the spectra obtained with each detector using the optimal filtering method~\cite{Gatti1986ProcessingTS}, the K$_\alpha$ 5.89 keV peak from the $^{55}$Fe source appears prominent and used for calibrating the detector response. The calibrated energy spectra are shown in Figure~\ref{Spectra_Au_Cu_Al/Au}. For the detectors with Au and Cu phonon collectors, the K$_\alpha$ and K$_\beta$ (6.49 keV) lines are not resolved, whereas the detector with the Al/Au phonon collector clearly resolves both lines. This is reflected quantitatively in the fitted K$_\alpha$ line width: the Au and Cu configurations yield FWHM values of \SI{1370.7(37.1)}{\electronvolt} and \SI{1282.3(19.6)}{\electronvolt}, while the Al/Au configuration has a narrower value of \SI{109.0(27.0)}{\electronvolt}. A shoulder on the low-energy side of the K$_\alpha$ peak is observed in the Al/Au configuration, which may be attributed to a contribution from the Au pad of the Al/Au phonon collector to the overall phonon collection efficiency. This hypothesis is currently under investigation through dedicated measurements.

The baseline resolution for absorber hits was evaluated following the procedure described in \cite{Angloher_2023}, and the results are summarized in Table~\ref{baselineresolution}. The detectors with the Au, Cu, and Al/Au phonon collectors achieved baseline resolutions of \SI{267.2(2.1)}{\electronvolt}, \SI{94.6(0.5)}{\electronvolt}, and \SI{21.5(0.3)}{\electronvolt}, respectively. A direct quantitative comparison of the performance across the three detector configurations is complicated by the fact that each detector was operated under different noise conditions, as shown in the right panel of Figure~\ref{Au_Cu_Al/Au_phonon_collector}.

\begin{table}[t]
\centering
\caption{Design parameters and baseline resolution of the three Si remoTES detector configurations studied in this work.}
\label{baselineresolution}
\setlength{\tabcolsep}{6pt}
\begin{tabular}{lccc}
\toprule
\multicolumn{4}{c}{\textbf{Common detector parameters}}\\
\midrule
Absorber volume (mm$^3$) & \multicolumn{3}{c}{$20 \times 10 \times 5$} \\
Au bonding wire diameter (\si{\micro\meter}) & \multicolumn{3}{c}{25 (ball bonded)} \\
TES design & \multicolumn{3}{c}{Au island} \\
\midrule
\multicolumn{4}{c}{\textbf{Detector-specific parameters}}\\
\midrule
 & \textbf{Au pc} & \textbf{Cu pc} & \textbf{Al/Au pc} \\
\midrule
Pad area (mm$^2$) & 3 & 3 & \makecell{2 (Al) \\ 0.13 (Au)} \\
Pad thickness (\si{\micro\meter}) & 0.20 & 0.25 & \makecell{1 (Al) \\ 0.6 (Au)} \\
Heat capacity (\si{\micro\joule\per\kelvin}) & $7.7\times10^{-7}$ & $1.5\times10^{-6}$ & $1.3\times10^{-7}$ (Au) \\
Baseline resolution & \SI{267.2(2.1)}{\electronvolt} & \SI{94.6(0.5)}{\electronvolt} & \SI{21.5(0.3)}{\electronvolt} \\
\bottomrule
\end{tabular}
\end{table}

Nevertheless, several conclusions can be drawn. When comparing the Au phonon collector to the Cu phonon collector, the Cu configuration achieves superior baseline resolution despite operating under slightly worse noise conditions, indicating an improvement due to the phonon collector. In contrast, when comparing the Al/Au to the Au and Cu phonon collectors, the Al/Au configuration exhibits better noise conditions. Nonetheless, the baseline resolution of $\sim$22 eV achieved with the Al/Au phonon collector has not been previously demonstrated with any remoTES detector. These results show that the choice of phonon collector material plays a significant role in the signal transmission and, consequently, in improving overall remoTES detector performance.
 
\section{Conclusions and outlook}
\label{sec:concl}
 
The remoTES design offers several key advantages, including compatibility with hygroscopic and fragile absorbers, scalable fabrication, and suitability for systematic optimization. This work investigated two design parameters, the Au port geometry and the phonon collector material, and their effect on signal transmission and consequently on the baseline resolution in Si-based remoTES detectors.
The Au island design, with the Au port deposited entirely on the W film, yields faster pulse decay than the Au bridge design, consistent with the increased thermal conductance dominating over the accompanying increase in W heat capacity. Comparing Au, Cu, and Al/Au phonon collectors, the latter achieved the best baseline resolution, \SI{21.5(0.3)}{\electronvolt}, compared to \SI{94.6(0.5)}{\electronvolt} for Cu and \SI{267.2(2.1)}{\electronvolt} for Au. This improvement is consistent with the much lower heat capacity of the Al/Au remoTES detector, its higher electron-phonon coupling, and the efficient Cooper-pair-breaking phonon absorption. Future work will focus on further optimizing the Al/Au phono collector geometry to improve the baseline resolution. Pulse shape studies of the detectors reported in this manuscript, using a BAT.jl~\cite{Schulz2021} fitting procedure developed for remoTES pulse shapes~\cite{Master_Sarah}, will be part of a future publication.
 
\backmatter
 
\bmhead{Acknowledgements}
We acknowledge the CRESST group at the Max-Planck Institute for Physics (MPP) for providing access to their facilities for the production of phonon collectors and the remoTES sensors used in this work. We also thank the MPP mechanical workshop for its invaluable technical support.
 
\section*{Declarations}
 
\bmhead{Funding}
This work was supported by the Klaus Tschira Foundation, the Austrian Science Fund FWF through the Sonderforschungsbereiche (SFB) (grant DOI: 10.55776/I6956), the FWF through the Scies4Free programme (grant DOI: 10.55776/DFH66), the FWF (grant DOI: 10.55776/PAT1239524), the Research Council of Finland, the Vilho, Yrj\"o, Kalle V\"ais\"al\"a Fund and Deutsche Forschungsgemeinschaft (DFG; German Research Foundation) under Germany’s Excellence Strategy - EXC 2094 - 390783311.
 
\bmhead{The authors declare no competing interests}
 
 
 
 
 
 
 
\bibliography{bibliography}
 
\end{document}